\documentstyle[preprint,aps,prl]{revtex}

\begin{document}

\baselineskip 1.2 cm

\newcommand{\IL}{({\bf v\cdot \hat{\sigma}})}
\newcommand{\WE}{{\bf v}}
\newcommand{\SI}{{\bf \hat{\sigma}}}
\newcommand{\ILL}{({\bf \hat{v}\cdot \hat{\sigma}})}
\newcommand{\UU}{{\bf u}}
\newcommand{\EN}{{\bf \hat{n}}}
\newcommand{\AC}{{\bf \hat{a}}}

\title{Lorentz's model with dissipative collisions}

\author{Philippe A. Martin}

\address{Institut de Physique Th\'eorique, Ecole Polytechnique F\'ed\'erale de
Lausanne, CH-1015, Lausanne, Switzerland}
 
\author{Jaros\l aw Piasecki}

\address{Institute of Theoretical Physics, University of Warsaw, Ho\.za 69, 
PL-00 681 Warsaw, Poland}

\date{\today}

\maketitle

\begin{abstract}
Propagation of a particle accelerated by an external field through a scattering 
medium is studied within the generalized Lorentz model allowing 
inelastic collisions. Energy losses at collisions are proportional
to $(1-\alpha^{2})$, where $0\le\alpha\le 1$ is the restitution coefficient.
For $\alpha =1$ (elastic collisions) there is no stationary state. It is
proved in one dimension that when $\alpha <1$  the stationary state exists .
The corresponding velocity distribution changes from a highly asymmetric
half-gaussian ($\alpha =0$) to an asymptotically symmetric distribution 
$\sim {\rm exp}[-(1-\alpha)v^{4}/2]$,  for $\alpha\to 1$. The identical scaling 
behavior in the limit of weak inelasticity is derived in three dimensions by a
self-consistent perturbation analysis, in accordance with the behavior of 
rigorously evaluated moments. The dependence on the external field scales out
in any dimension, predicting in particular the stationary current to be 
proportional to the square root of the external acceleration.
\end{abstract}
\vspace{5mm}

\section{INTRODUCTION}

The object of the present paper is to study the propagation of a particle 
through a medium composed of immobile spherically symmetric scatterers (of 
infinite mass), randomly distributed in space with some number density $n$. 
Between collisions the particle moves with acceleration ${\bf a}$, acted upon by
a constant and uniform external field. At binary collisions its velocity 
${\bf v}$ is instantaneously transformed according to the law
\begin{equation}
{\bf v}\;\; \rightarrow \;\; {\bf v' =  v} - (1+\alpha){\bf 
(v\cdot \hat{\sigma})} \hat{\sigma}\label{1}
\end{equation}
where $0\leq \alpha \leq 1$ is the so called restitution coefficient, 
and $\hat{\sigma}$ denotes the unit vector along the line passing through the 
centers of colliding particles at the moment of impact. Whereas the projection
of velocity ${\bf v}$ on the direction tangent to the surface of the scatterer
at the point of contact is not changed, the component $({\bf v\cdot \hat{\sigma}
})$ along the normal gets multiplied by $(-\alpha )$. In the extreme case of
$\alpha =0$, the postcollisional velocity reduces to the tangential component.

When $\alpha =1$, collisions (\ref{1}) are perfectly elastic, and we recover 
the situation of the classical Lorentz model of electric
conductivity in metals \cite{lorentz05}. It can be proved that there is no 
stationary state in this case \cite{pias7993},\cite{olaussen82}. 
This is because the moving particle does not loose energy 
at collisions, and gets asymptotically heated by the field beyond any bounds. 
In particular, its mean kinetic energy diverges as $\sim t^{2/3}$ when the 
time $t\to \infty$.

The aim of this work is to show, that in the case of inelastic collisions 
(\ref{1}) the stationary state becomes possible in the whole range of 
$0\leq \alpha <1$. Once $\alpha <1$, there occurs dissipation of the kinetic
energy $E=mv^{2}/2$ at encounters, as in accordance with (\ref{1}) $E$ suffers 
the transformation
\begin{equation}
E \;\; \rightarrow \;\; E' = E - m(1-\alpha^{2})(v\cdot \hat{\sigma})^{2}/2
\label{2}
\end{equation}
The inelastic dissipation (\ref{2}) turns out to suffice to balance the energy
flow from the external field.

The basis for the subsequent analysis is the linear Boltzmann equation
satisfied by the probability density $f({\bf v};t)$ for finding the
propagating particle with velocity  ${\bf v}$ at time $t$. In writing this
equation
one must take into account that the jacobian of transformation (\ref{1}) equals
$\alpha$ (dissipative collisions are contracting the volume in the velocity 
space), and that the inverse transformation is obtained by replacing $\alpha $ 
in (\ref{1}) by $\alpha^{-1}$. The kinetic equation reads
\begin{equation}
\left( \frac{\partial}{\partial t} + {\bf a}\cdot\frac{\partial}{\partial \WE}
 \right)f(\WE;t) = \frac{|\WE|}{\lambda \pi} \int d\SI \ILL \theta \ILL
 \left\{ \alpha^{-2}f[\WE -(1+\alpha^{-1}\IL \SI );t]
 - f(\WE;t) \right\} \label{3}
\end{equation}
Here $\lambda = (\pi nR^{2})^{-1}$ denotes the mean free path ($R$ is the sum
of the particle and the scatterer radii), ${\bf \hat{v}}$ is the unit velocity
vector, and the integration with respect 
to $d\SI$ spreads over a unit sphere (solid angle). In the gain term (due to
collisions) the factor $\alpha^{-2}$ compensates for the
contraction of the velocity space and gives the proper value to the
precollisional velocity of approach  $\IL /\alpha$. Owing to this factor
the velocity integral of the collision term vanishes, which permits to deduce 
from (\ref{3}) the continuity equation. 

A convenient equivalent form of equation (\ref{3}) reads  
\begin{equation}
\left( \frac{\partial}{\partial t} + {\bf a}\cdot\frac{\partial}{\partial \WE}
 \right)f(\WE;t) = \frac{|\WE|}{\lambda }\left\{ \alpha^{-2}\int 
 \frac{d\EN}{4\pi } f\left[\frac{1+\alpha }{2\alpha }|\WE|\EN -
 \frac{1-\alpha }{2\alpha }\WE ;t\right]
 - f(\WE;t) \right\} \label{4}
\end{equation}
It can be obtained from (\ref{3}) by changing the angular integration 
variables from $d\SI = sin\psi d\psi d\phi$ (where $cos\psi =\ILL$) to 
$d\EN = sin\chi d\chi d\phi$, with $\chi = 2\psi$. When $\alpha =1$, we recover 
the kinetic equation derived originally by Lorentz \cite{lorentz05},   

Our work is closely related to the study of spontaneous percolation in three
dimensions of Wilkinson and Edwards \cite{wilkinson82}. The dynamics of particles 
falling under gravity through a fixed scattering medium was analyzed therein  on the 
basis of a Boltzmann-like equation with a phenomenological scattering function. 
The perturbative methods developed in \cite{wilkinson82} permitted to analyze
the low inelasticity limit leading to the same qualitative predictions as those
derived here. The novelty of our contribution consists mainly in proving by an
explicit construction the existence
of a stationary state in one dimension, and in developing an original  
self-consistent perturbative expansion in three dimensions, inspired by the
scaling structure appearing in one dimension. It is to be stressed that the 
rigorous one-dimensional results firmly support the predictions of the perturbative
analysis in three dimensions, as the properties of the system in both cases
are qualitatively the same. We could also calculate rigorously some of the
moments of the stationary distribution in three dimensions. Their asymptotic
behavior in the elastic limit coincides with that predicted by the proposed
perturbation scheme.
In Section II the dimensional analysis of the stationary solution to the
kinetic equation (\ref{4}) is performed. Then, in Section III 
the one-dimensional version of the model is considered. 
We prove therein the existence of a stationary solution and we
analyze its dependence on parameter $\alpha$.  Section IV is devoted to the 
extension of our results to three dimensions. This is obtained in the low 
inelasticity limit $\alpha \to 1$ by a self-consistent construction of a 
stationary state with appropriate scaling behavior. The paper ends with 
concluding comments. 

\section{DIMENSIONAL ANALYSIS}

It is as usual instructive to formulate the theory in terms of dimensionless 
variables. We thus put
\begin{equation}
{\bf v} = \sqrt{a\lambda}{\bf u} \label{5} 
\end{equation}
where $a=|{\bf a}|$. Denoting by $F(\UU)$ the stationary dimensionless 
distribution we find from (\ref{4}) the equation
\begin{equation}
\AC \cdot \frac{\partial}{\partial \UU}F(\UU) = |\UU|\left\{  \alpha^{-2}\int 
 \frac{d\EN}{4\pi } F\left[\frac{1+\alpha }{2\alpha }|\UU|\EN -
 \frac{1-\alpha }{2\alpha }\UU \right] - F(\UU) \right\}  \label{6}
\end{equation}
 The amplitude $a$ of the external field disappeared from
equation (\ref{6}). Only the unit vector $\AC$ shows there, indicating the
direction of acceleration. Hence, the field dependence of the moments of
distribution $F$ can be obtained by the velocity scaling (\ref{5}). In 
particular
\begin{equation}
< \WE > = \sqrt{a\lambda} < \UU > \sim \sqrt{a},\;\;\; < v^{2} > \sim a
\label{7}
\end{equation}

The fact that the stationary particle current is proportional to the square
root of the field is of no surprise. The immobile scatterers represent a
zero-temperature medium, and the only energy scale comes from the acceleration
$a$. The relations (\ref{7}) could be thus predicted on purely dimensional
grounds. At fixed restitution coefficient $\alpha$, one cannot expect
the linear response even for very weak fields.

The above conclusions apply in any dimension. In the study of the stationary 
state we shall use in the coming sections the dimensionless form  (\ref{6}) 
of the kinetic equation.

\section{ONE-DIMENSIONAL STATIONARY VELOCITY DISTRIBUTION}

In one dimension equation (\ref{6}) takes the form 
\begin{equation}
\frac{d}{du}F(u) = |u|\left[ \alpha^{-2}F(-\alpha^{-1}u)
- F(u)\right] \label{8}
\end{equation}
Putting
\begin{equation} 
F(u)\equiv G(u|u|/2) \label{80}
\end{equation}
we find
\begin{equation}
G'(s) + G(s) = \alpha^{-2}G(-\alpha^{-1}s), \label{9}
\end{equation}
where $G'$ denotes the derivative of $G$. 
Applying to (\ref{9}) the Fourier transformation one finds
\begin{equation}
\hat{G} (k) = (1+ik)^{-1}\hat{G} (-\alpha^{2}k) = \hat{G}(0)
\prod_{r=0}^{\infty}[1+i(-\alpha^{2})^{r}k]^{-1} \label{10}
\end{equation}
where\[ \hat{G} (k) = \int ds e^{-iks}G(s) \] 

For $0\le \alpha <1$ the infinite product converges which proves the existence
of the stationary state. This is an important conclusion indeed, showing the
fundamental difference with respect to the Lorentz model with elastic 
collisions. Clearly, an arbitrary degree of inelasticity suffices to balance the
energy flow from the external field.

A particularly simple result is found for perfectly inelastic collisions. 
Indeed, when $\alpha =0$ equation (\ref{10}) reduces to
\begin{equation}
\hat{G}(k) = \frac{\hat{G}(0)}{(1+ik)}  \label{11}
\end{equation}

Inverting the Fourier transformation and using (\ref{80}) we find
\begin{equation}
F(u) = \theta(u) \sqrt{\frac{2}{\pi}}{\rm exp}(-u^{2}/2)  \label{12}
\end{equation}
In one dimension each collision with $\alpha =0$ completely stops the particle, 
dissipating the whole energy absorbed from the field. 
So, in a stationary flow the velocities oriented against
the field are not possible (the $\theta $ factor in (\ref{12})). It is quite
remarkable that the half space of possible velocities gets a gaussian weight.

Let us turn now to equation (\ref{9}). It is equivalent  to the system of two 
coupled equations 
\begin{equation}
G'_{+}(s) + G_{-}(s) = -\alpha^{-2} G_{-}(s/\alpha^{2}) \label{13}
\end{equation}
\[ G'_{-}(s) + G_{+}(s) = \alpha^{-2} G_{+}(s/\alpha^{2}) \]
where $G_{+}(s)= [G(s)+G(-s)]/2$ and $G_{-}(s)= [G(s)-G(-s)]/2$ are the 
symmetric and antisymmetric parts of $G$, respectively. The second relation in
(\ref{13}) can be solved for $G_{-}(s)$ yielding
\begin{equation}
G_{-}(s) = \int_{s}^{s\alpha^{-2}} dz\,G_{+}(z) \label{14}
\end{equation}
We then get from (\ref{13}) a closed equation for the symmetric part
\begin{equation}
G'_{+}(s) = -\int_{s}^{s\alpha^{-2}} dz\,G_{+}(z)
-\alpha^{-2}\int_{s\alpha^{-2}}^{s\alpha^{-4}} dz\,G_{+}(z) \label{15}
\end{equation}
In order to study the low inelasticity limit we define a small parameter
$\epsilon = 1-\alpha $. When $\epsilon \ll 1$, equation (\ref{15}) takes the
simple asymptotic form $G'_{+}(s)= -4\epsilon sG_{+}(s)$, and we find
\begin{equation}
G_{+}(s) = C\epsilon^{1/4}{\rm exp}(-2\epsilon s^{2}) \label{16}
\end{equation}
Using then the relations (\ref{14}) and (\ref{80}) we can determine the 
asymptotic form of the velocity distribution for weakly inelastic scattering. 
It reads
\begin{equation}
F(u) = \frac{C\epsilon^{1/4}}{\sqrt{2}}(1+\epsilon u|u|){\rm exp}(-\epsilon
u^{4}/2), \;\;\;\; \epsilon \ll 1 \label{17}
\end{equation}
The value of constant $C$ is independent of $\epsilon$ and follows form 
the normalization condition
\[ C^{-1} = \int dw {\rm exp}(-2w^{4}) \]
The dominant term in (\ref{17}) is symmetric and depends on the
scaled variable $w=\epsilon u^{4}$. It implies the divergence  $\sim \epsilon^
{-1/2}$ of the kinetic energy $<u^{2}>$ for $\epsilon \to 0$. This divergence
reflects the disappearance of the stationary state in the case of elastic
collisions. 
In fact, the kinetic equation (\ref{8}) permits to calculate directly some of 
the moments of distribution $F$, and to predict divergences in the elastic 
limit. An interesting example is provided by the recurrence relation
\begin{equation}
<u^{j}|u|^{j}> = \frac{2j-1}{1-(-1)^{j}\alpha^{2j-1}}<u^{j-1}|u|^{j-1}>,
\;\;{\rm with} \;\;<u|u|>=(1+\alpha)^{-1}
\label{18}
\end{equation} 
which follows directly from (\ref{8}) when mutiplied by $u^{j}|u|^{j-1}$ and 
integrated over the velocity space. For $j=2$, we get from (\ref{18}) 
\[ <u^{4}> = 3[(1+\alpha)(1-\alpha^{3})]^{-1} ,\]
implying the divergence $\sim \epsilon^{-1}$ of the fourth moment. It should
be stressed that the
asymptotic state (\ref{17}) is consistent with the exact relations (\ref{18}).
This result can also be obtained directly from the explicit solution by studying
the asymptotics of the infinite product in (\ref{10}) as $\alpha\to 1$.
The high kinetic energy for $\epsilon \to 0$ implies a large collision 
frequency, which makes the asymptotic distribution (\ref{17}) almost symmetric.
In this situation the stationary current is very weak. Indeed, we find
\begin{equation}
<u> = C\,\epsilon^{1/4}/\sqrt{2}
\end{equation}

Before closing let us note that the above results can be generalized to the
case where the collision frequency is proportional to $|u|^{\gamma},\;\;
0\le \gamma <1$, rather than to $|u|$. The corresponding generalization of the
kinetic equation (\ref{8}) reads
\begin{equation}
F'(u) = |u|^{\gamma}\left[ \alpha^{-(1+\gamma)}F(-\alpha^{-1}u)
- F(u)\right] \label{19}
\end{equation}
and the substitution \[F(u) \equiv G[u|u|^{\gamma}/(\gamma +1)] \]
transforms (\ref{19}) into
\begin{equation}
G'(s) = \alpha^{-(1+\gamma )}G(-s\alpha^{-(1+\gamma )}) - G(s)
\end{equation}
Performing then the same analysis as for $\gamma =1$, we find the generalization
of the asymptotic formula (\ref{17})
\begin{equation}
F(u) = C\epsilon^{1/2(\gamma +1)} (\gamma +1)^{-1/\gamma +1}
(1+\epsilon u|u|^{\gamma}){\rm exp}[-\epsilon u^{2(\gamma +1)}/(\gamma +1)], 
\;\;\;\; \epsilon \ll 1 \label{20}
\end{equation}

Again the dominant term in (\ref{20}) is symmetric, and the first
two moments follow the asymptotic law
\[ <u> \sim \epsilon^{\gamma/2(\gamma +1)},\;\;\;
 <u^{2}> \sim \epsilon^{-1/\gamma +1}\]
An interesting case is that of the so called Maxwell gas, where $\gamma =0$,
so that the collision frequency becomes independent of the velocity. The 
stationary current stays then finite even in the limit $\epsilon \to 0$. 
However, the $\sim \epsilon^{-1}$
divergence in the kinetic energy indicates the infinite absorption of energy
from the  field for $\epsilon =0$.

Our analysis for  $\gamma =1$ (free motion between collisions) shows that the 
stationary velocity  distribution changes from a highly asymmetric 
half-gaussian (\ref{12}) at the strongest dissipation, to an almost symmetric 
scaled distribution  $\sim {\rm exp}(-\epsilon u^{4})$ in
the weak dissipation limit. The experience from the studies of the classical
Lorentz model showed the independence from the spatial dimension in the 
qualitative behavior. We can thus expect the appearance of the same scaling 
structure in three dimensions.

\section{STATIONARY VELOCITY DISTRIBUTION IN THREE DIMENSIONS}

In three dimensions the stationary velocity distribution $F(\UU)$ satisfies the 
equation (see (\ref{4}))

\begin{equation}
\AC \cdot\frac{\partial}{\partial \UU}F(\UU) = |\UU|\left\{ \alpha^{-2}\int 
 \frac{d\EN}{4\pi } F\left[\frac{1+\alpha }{2\alpha }|\UU|\EN -
 \frac{1-\alpha }{2\alpha }\UU \right] - F(\UU) \right\} \label{21}
\end{equation}

Multiplying (\ref{21}) by a function $\psi(\UU)$ and integrating over the 
velocity space one obtains the dual equation of the form
\begin{equation}
<\left\{ \AC \cdot \frac{\partial}{\partial \UU}\psi(\UU)
+ |\UU|\int \frac{d\EN}{4\pi } \psi \left[ \frac{1-\alpha }{2}\UU 
- \frac{1+\alpha }{2}|\UU|\EN \right] - |\UU|\psi(\UU) \right\} >\;\; = \;\; 0 
\label{22}
\end{equation}
Here $<...>$ denotes the average with respect to the probability density $F$.
One can obtain from (\ref{22}) useful information by an appropriate choice of 
$\psi$. In fact, some moments of $F$ can be evaluated exactly in this way for 
any value of $\alpha$. So, for instance
putting $\psi (\UU) = \UU \cdot \AC $ yields the formula

\begin{equation}
(1+\alpha)< |\UU|(\UU \cdot \AC ) > = 2 \label{23}
\end{equation}

And when  $\psi = |\UU|^{3} $, we find
\begin{equation}
3< |\UU|(\UU \cdot \AC ) > = \left[ 1-\frac{2(1-\alpha^{5})}{5(1-\alpha^{2})}
\right]<|\UU|^{4}> \label{24}
\end{equation}
Combining (\ref{23}) and (\ref{24}) permits to determine the fourth moment
\begin{equation}
(1-\alpha)(3+6\alpha +4\alpha^{2}+2\alpha^{3})<|\UU|^{4}> = 30 \label{25}
\end{equation}
In accordance with the remark made at the end of section III, we find here the
same divergence $<|\UU|^{4}>\sim \epsilon^{-1}$ for $\epsilon =(1-\alpha) \to 0$ as in
one dimension. Let us finally note the relation
\begin{equation}
4<(\UU \cdot \AC )> = <|\UU |^{3}>(1-\alpha^{2}) \label{26}
\end{equation}
following from (\ref{22}) for $\psi (\UU)= |\UU|^{2}$.

In order to determine the distribution $F$ in the limit of weak inelasticity we
put $\alpha =(1-\epsilon)$ in equation (\ref{21}), and expand the inelastic 
collision law up to terms linear in $\epsilon$. 
Clearly, the stationary state can depend on two variables only, 
$u=|\UU|$ and $ \mu = \AC \cdot \hat{\UU}$. Expressing the differential operator 
in (\ref{21}) in terms of $u$ and $\mu$ one finds eventually the following 
asymptotic equation
\begin{equation}
\left( \mu\frac{\partial}{\partial u}+\frac{1-\mu^{2}}{u}
 \frac{\partial}{\partial \mu}+u\right)F_{A}(u,\mu) = -\mu \frac{d}{du}F_{S}(u)
 +\epsilon \left( \frac{1}{2u^{2}}\frac{d}{du}[u^{4}F_{S}(u)] 
 - \frac{\mu}{4}\frac{d}{du}[u^{2}
 \int_{-1}^{1}d\sigma\,\sigma F_{A}(u,\sigma)]\right) \label{27}
\end{equation} 

By analogy with the approach developed in the one-dimensional case,
we introduced here the spherically symmetric projection of $F$
\[ F_{S}(u) = \int \frac{d\hat{\UU}}{4\pi}F(\UU)     \]
and the deviation form spherical symmetry $F_{A}=F-F_{S}$.

An important relation between $F_{S}$ and $F_{A}$ is obtained by integration of
(\ref{27}) over the whole range of the angular variable $-1\le \mu \le 1$. 
One finds
\begin{equation}
\int_{-1}^{1}d\mu\,\mu F_{A}(u,\mu) = \epsilon u^{2}F_{S}(u) \label{28}
\end{equation}

In order to proceed, we shall assume that for $\epsilon \ll 1$ the state $F$ 
depends on velocity via a scaled variable $\epsilon^{x}u$. The exact relations 
(\ref{23})-(\ref{26}) imply then the  value $x=1/4$. Indeed, we know from 
(\ref{25}) that the fourth moment diverges like $\epsilon^{-1}$, whereas the 
assumed scaling predicts the $\sim \epsilon^{-4x}$ behavior. 
So, taking also into account relation (\ref{28}), we write the perturbative 
expansions of $F_{S}$ and $F_{A}$ as
\begin{equation}
F_{S}^{\epsilon}(w) = F_{S}^{0}(w)+\sqrt{\epsilon}F_{S}^{1}(w)+
\epsilon F_{S}^{2}(w)+ ... \label{29}
\end{equation}
\[ F_{A}^{\epsilon}(w,\mu) = \sqrt{\epsilon}F_{A}^{1}(w,\mu)+
\epsilon F_{A}^{2}(w,\mu)+ ... , \]
where $w \equiv \epsilon^{1/4}u$. 

The expansion of the non-spherical part $F_{A}$ does not contain the zero order
term in accordance with the relation (\ref{28}) between $F_{S}$ and $F_{A}$.
Inserting expansions (\ref{29}) into equations (\ref{27}) and  (\ref{28}) to
lowest order in $\epsilon $ we find
\begin{equation}
wF_{A}^{1}(w,\mu) = - \mu\frac{d}{dw}F_{S}^{0}(w),\;\;\;\;\;\;
\int_{-1}^{1}d\mu \mu F_{A}^{1}(w,\mu) = w^{2}F_{S}^{0}(w) \label{30}
\end{equation}

From (\ref{30}) there follows a closed equation for $F_{S}^{0}$
\[ 3w^{3}F_{S}^{0}(w) = -2 \frac{d}{dw}F_{S}^{0}(w) \]
The lowest order terms in expansion (\ref{29}) are then readily 
determined. The asymptotic formula for the stationary distribution reads
\begin{equation}
F(w,\mu) = F_{S}^{0}(w)+\sqrt{\epsilon}F_{A}^{1}(w,\mu)= 
C\epsilon^{3/4}(1+3\sqrt{\epsilon}\mu w^{2}/2){\rm exp}(-3w^{4}/8), \label{31}
\end{equation}
where the constant $C$ assures the normalization. Inserting into (\ref{31}) 
$w=\epsilon^{1/4}u$ and $\mu =\AC \cdot\hat{\UU}$, we obtain the same 
structure of the state $F$ as in one dimension (compare with (\ref{17})). 
The same $\epsilon$-dependence of the moments of $F$ thus also holds.

This perturbative analysis can be continued. We checked that at the next step
the following relations are found
\[F_{S}^{1}(w) \equiv 0,\]
\[ F_{A}^{2}(w,\mu) = P_{2}(\mu)[1+w\frac{d}{dw}]F_{S}^{0}(w), \]
where $P_{2}(\mu)=(1-3\mu^{2})/2$ is the second Legendre polynomial. This 
confirms self-consistency of expansion (\ref{29}) in powers of 
$\sqrt{\epsilon}$. 

\section{CONCLUDING COMMENTS}

We proved the existence of a stationary state in the case of a one-dimensional
propagation of a uniformly accelerated particle suffering inelastic collisions
with randomly distributed scatterers. The velocity distribution, strongly
asymmetric at high dissipation, showed the scaling behavior in the elastic
limit $\epsilon =1-\alpha \to 0$, recovering asymptotically the spherical 
symmetry $F \sim {\rm exp(-\epsilon u^{4})}$. 

The same scaling properties could be derived in three dimensions  
by a suitable perturbation method. Of course, it would be desirable to provide 
a rigorous existence proof also in three dimensions. 

In the one-dimensional case we could determine the form of the velocity 
distribution for perfectly inelastic collisions (see (\ref{12})). In three
dimensions, passing in the dual equation (\ref{22}) to limit 
$\alpha \to 0$, one can deduce in a straightforward way the form of the 
equation satisfied by the state $F$ at $\alpha =0$
\begin{equation}
\pi \AC \cdot\frac{\partial}{\partial \UU}F(\UU) = 
\int d{\bf w}\,\delta({\bf w}\cdot\UU)F({\bf w}+\UU) - \pi|\UU|F(\UU) \label{32}
\end{equation}

It should be realized that in contradistiction to the one-dimensional case
the particle is not stopped at encounters with $\alpha =0$, but
its postcollisional velocity reduces then to the component tangent to the 
surface of the scatterer. It is thus clear that the solution to (\ref{32}) 
can give non-zero probability weight only when $\AC \cdot \UU > 0$. Solving
equation (\ref{32}) is another open problem related to our study.

Finally, an interesting question would be to explore the dynamics of approach 
to the stationary distribution. This problem has been solved only in
one dimension for $\alpha =0$ \cite{piasecki83}.

\section*{ACKNOWLEDGMENTS}

J.Piasecki acknowledges the hospitality at the Institute of Theoretical Physics
of the Ecole Polytechnique F\'ed\'erale de Lausanne (Switzerland), and financial
support by the KBN (Committee for scientific Research, Poland), 
grant 2 P03B 03512.

\end{document}